\documentclass[aps,prb,twocolumn,superscriptaddress,floatfix]{revtex4}
\usepackage{graphicx}
\usepackage{psfrag}
\usepackage{color}
\usepackage{subfigure}
\usepackage{amsmath}
\definecolor{dred}{rgb}{0.7,0.0,0.0}
\allowdisplaybreaks 
 
\begin{document}

\title{Optical Conductivity Anisotropy in the \\
Undoped Three-Orbital Hubbard Model for the Pnictides}

\author{Xiaotian Zhang}
\author{Elbio Dagotto} 

\affiliation{Department of Physics and Astronomy, The University of
  Tennessee, Knoxville, TN 37996} 
\affiliation{Materials Science and Technology Division, Oak Ridge
  National Laboratory, Oak Ridge, TN 32831}

\date{\today}

\begin{abstract}
The resistivity anisotropy unveiled
in the study of detwinned single crystals of the undoped 122 pnictides
is here studied using the two-dimensional three-orbital Hubbard model in the mean-field
approximation. Calculating the Drude weight in the $x$ and $y$ directions 
at zero temperature for a ${\bf Q}$=($\pi$,0) magnetically ordered state, the
conductance along the antiferromagnetic direction is shown to be 
larger than along the ferromagnetic direction. This effect is caused 
by the suppression of the $d_{yz}$ orbital at the Fermi surface, but additional
insight based on the momentum dependence of the transitions induced 
by the current operator is provided. It is shown that the effective suppression of the 
inter-orbital hopping $d_{xy}$ and $d_{yz}$ along the $y$ direction is the main cause of
the anisotropy.


\end{abstract}

\maketitle


{\it Introduction.} 
One of the most intriguing puzzles in the study of the
Fe-based high temperature superconductors~\cite{johnston}
is the discovery of unexpected transport anisotropies
in detwinned single crystals of doped and undoped AFe$_2$Sr$_2$ (A = Ba, Sr, Ca).\cite{aniso}
Studies of the in-plane resistivity~\cite{chu}
showed that the effect is the largest at low doping
$x$$\sim$2-4\% in Ba(Fe$_{1-x}$Co$_x$)$_2$As$_2$,
but it is present even in the undoped limit $x$=0 at low temperatures, i.e.
in the magnetically ordered state with wavevector ($\pi$,0). 
Recent studies~\cite{blomberg}
for the undoped 122 materials have revealed
 a low-temperature anisotropy (defined as $R = \rho_b/\rho_a -1$) 
$R$$\sim$0.4, 0.35, and 0.09, for A = Ba, Sr, and Ca, respectively.
This anisotropy is counter-intuitive because along the $a$-axis the spins order in an
antiferromagnetic (AFM) arrangement, while along the $b$-axis they are ferromagnetic (FM).
Intuition based on, e.g., double-exchange mechanisms for manganites would suggest that
the FM direction should be less resistive than the AFM one. Optical conductivity measurements
concluded that this unexpected anisotropy is caused by changes 
in  the populations of the orbitals $d_{xz}$ and $d_{yz}$ at the Fermi surface (FS),\cite{dusza} 
in agreement with early mean-field studies where this
unbalanced FS orbital population, without long-range
orbital order, led to results compatible with photoemission techniques.\cite{owr}

Several calculations have recently addressed the experimentally observed transport anisotropy.
Using a five-orbital Hubbard model treated in a mean-field approximation, 
and calculating the Drude weights
via the Fermi velocities at the FS, results compatible 
with experiments were reported.~\cite{valenzuela} This agreement was 
observed in regimes where long-range orbital
order is not present, and indeed the FS redistribution of spectral weight among the
$d_{xz}$ and $d_{yz}$ orbitals caused by the
($\pi$,0) magnetic order\cite{owr} is needed  to understand the experimental results.
Other calculations also for the five-orbital Hubbard model arrived to 
similar conclusions.\cite{tohyama,valenti}

In this publication, the transport anisotropy found in experiments
is revisited from the perspective of a simpler three-orbital Hubbard model.\cite{three}
Our goal is to refine the intuitive 
explanations given in Refs.~\onlinecite{valenzuela,tohyama}, by focusing 
on the three orbitals widely believed to be the most important in pnictides, namely
$d_{xz}$, $d_{yz}$, and $d_{xy}$, and also by identifying the electronic 
hopping amplitudes that cause the anisotropy. Our main results are that the experimentally
observed anisotropy clearly appears in the three-orbital model, in a state that is
($\pi$,0) magnetically ordered, and it is mainly caused by the suppression of
{\it inter}-hopping $d_{xz}$-$d_{yz}$ processes along the FM direction.

{\it Models and methods.} In this manuscript, the three-orbital Hubbard model
for the pnictides at overall electronic density $n$=4/3 (per site and per orbital) 
will be used.\cite{three} The hopping amplitudes that reproduce
the FS in the paramagnetic state, with hole and electron pockets, were already
provided and discussed in detail in Ref.~\onlinecite{three}. However, to help the readers in the
understanding of our results, in Table~\ref{tab:hopp3} 
these intra- and inter-hopping amplitudes (in eV units)
are provided again. From Table~\ref{tab:hopp3} 
note that the hoppings involving the $d_{xy}$
orbital, both intra-orbital and also inter-orbital with $d_{xz}$ and $d_{yz}$, are the
largest in value, inducing a large Fermi velocity in the regions  of the FS where the
$d_{xy}$ orbital dominates. This suggests that the $d_{xy}$ may play an important role in the anisotropy.
The full three-orbital Hubbard model also contains a Coulombic on-site interaction
(see Eq.~(11) of Ref.~\onlinecite{three}), involving an intra-orbital 
Hubbard repulsion $U$, an inter-orbital repulsion $U'$, and a Hund coupling $J$, with
the constraint $U'$=$U$-$2J$. The  Hartree mean-field approximation
used here has also been much discussed in previous literature and the reader is
referred to Refs.~\onlinecite{three,rong,luo} for details. The mean-field order parameters
are the three electronic densities of each orbital, i.e. 
$n_{xz}$, $n_{yz}$, and $n_{xy}$,
and the three magnetic moments $m_{xz}$, $m_{yz}$, and $m_{xy}$, and they are all determined
via the minimization of the Hartree mean-field energy. In the mean-field equations,
the wavevector ${\bf Q}$ = ($\pi$,0) is assumed.

\begin{table}
\centering
 \begin{tabular}{|c|cccc|}\hline
$t^{\alpha\beta}_{\bf{il}}$ & $\bf{l}=x$   & $\bf{l}=y$   & $\bf{l}=x+y$    & $\bf{l}=x-y$\\
\hline
  $\alpha\beta=11$                  &$-0.06$       &$-0.02$       &$-0.03$           &$-0.03$   \\
\hline
  $\alpha\beta=22$                  &$-0.02$       &$-0.06$       &$-0.03$           &$-0.03$   \\
\hline
  $\alpha\beta=33$                  &$0.2$         &$0.2$         &$-0.3$            &$-0.3$ \\
\hline
  $\alpha\beta=12$                  &$0.0$         &$0.0$         &$-0.01$          &$0.01$        \\
\hline
  $\alpha\beta=13$                  &$0.2$         &$0.0$         &$-0.1$           &$-0.1$       \\
\hline
  $\alpha\beta=23$                  &$0.0$         &$0.2$        &$-0.1$             &$0.1$       \\
\hline
 \end{tabular}
\caption{Tight-binding (TB) hopping parameters of the three-orbital Hubbard model
  used in this manuscript. 
The energy unit is eV. The labeling convention is $1$=$d_{xz}$, $2$=$d_{yz}$, $3$=$d_{xy}$. 
The 13 and 23 hoppings are all affected by a factor $(-1)^{|i|}=(-1)^{{i_{x}}+{i_{y}}}$, with ${\bf i}$=($i_x$,$i_y$) being
the label of the Fe sites of a two-dimensional lattice. This modulation takes into account the
two-Fe unit cell of the original FeAs layers.\cite{three} The TB Hamiltonian is defined as 
$H_{\rm HT}$=$\sum_{{\bf i}{\bf l}\alpha \beta \sigma} t^{\alpha \beta}_{{\bf i}{\bf l}}
(c^\dagger_{{\bf{i}},\alpha,\sigma}  c_{{\bf{i+l}},\beta,\sigma}  + h.c.)$, where  
$c^\dagger_{{\bf{i}},\alpha,\sigma}$ creates an electron at orbital $\alpha$
of  site ${\bf i}$ with spin projection $\sigma$. ${\bf i+l}$ denotes nearest and
next-nearest neighbor sites to ${\bf i}$.
\label{tab:hopp3}} 
\end{table}

Let us focus now on the optical conductivity $\sigma(\omega)$.
Following well-known 
computational studies of $\sigma(\omega)$ in the context of the cuprates,\cite{Elbio} 
let us define first the paramagnetic current operators in the two
directions as 

\begin{equation}\begin{split}\label{E.Htb1}
 {\hat{j}}_{x}&=\sum_{\langle \mathbf{i,l=\hat{x},\hat{x}+\hat{y},\hat{x}-\hat{y}} \rangle} \sum_{\alpha,\beta,\sigma}
-it^{\alpha\beta}_{\mathbf{il}} (c^\dagger_{\mathbf{i},\alpha,\sigma}
c^{}_{\mathbf{i+l},\beta,\sigma} - h.c.),\\
 {\hat{j}}_{y}&=\sum_{\langle \mathbf{i,l=\hat{y},\hat{x}+\hat{y},-\hat{x}+\hat{y}} \rangle} \sum_{\alpha,\beta,\sigma}
-it^{\alpha\beta}_{\mathbf{il}} (c^\dagger_{\mathbf{i},\alpha,\sigma}
c^{}_{\mathbf{i+l},\beta,\sigma} - h.c.),\\
\end{split}\end{equation} 

\noindent while the kinetic energy operators are

\begin{equation}\begin{split}\label{E.Htb2}
 {\hat{T}}_{x}&=\sum_{\langle \mathbf{i,l=\hat{x},\hat{x}+\hat{y},\hat{x}-\hat{y}} \rangle} \sum_{\alpha,\beta,\sigma}
t^{\alpha\beta}_{\mathbf{il}} (c^\dagger_{\mathbf{i},\alpha,\sigma}
c^{}_{\mathbf{i+l},\beta,\sigma} + h.c.),\\
  {\hat{T}}_{y}&=\sum_{\langle \mathbf{i,l=\hat{y},\hat{x}+\hat{y},-\hat{x}+\hat{y}} \rangle} \sum_{\alpha,\beta,\sigma}
t^{\alpha\beta}_{\mathbf{il}} (c^\dagger_{\mathbf{i},\alpha,\sigma}
c^{}_{\mathbf{i+l},\beta,\sigma} + h.c.).\\
\end{split}\end{equation}

The total current, up to the first order term in the external field ${\bf{A}}$=($A_x$,$A_y$), 
can be written as ${\hat{J}}_{x}$=$({\hat{j}}_{x}+{\hat{T}}_{x}A_{x})/N$ and ${\hat{J}}_{y}$=$({\hat{j}}_{y}+{\hat{T}}_{y}A_{y})/N$.
The real part of the optical conductivity in the $x$ direction is given by:~\cite{Elbio}

\begin{equation}\begin{split}
 Re \sigma_{xx}(\omega)&=D_{x}\delta(\omega)\\
                       &+\frac{\pi}{N}\sum_{n\not=0}
    \frac{|\langle \phi_{0}|{\hat{j}}_{x}|\phi_{n} \rangle| ^{2}}{E_{n}-E_{0}}\delta(\omega-(E_{n}-E_{0})),\\
\end{split}\end{equation} 

\noindent
and from the $\sigma(\omega)$ sum-rule, it can be shown that 
the Drude weight in the $x$ direction $D_{x}$ is~\cite{Elbio}

\begin{equation}\begin{split}
 \frac{D_{x}}{2\pi}&=\frac{\langle \phi_{0}|-{\hat{T}}_{x}|\phi_{0} \rangle}{2N}-\frac{1}{N}\sum_{n\not=0}
    \frac{|\langle \phi_{0}|{\hat{j}}_{x}|\phi_{n} \rangle| ^{2}}{E_{n}-E_{0}},\\
\end{split}\end{equation}

\noindent
where $\phi_{0}$ is the many-body ground state 
(in this case the mean-field ${\bf Q}$=($\pi$,0) state),
and $\phi_{n}$ represents
the many-body excited states, also produced in the mean-field calculation, 
with $E_{0}$ and $E_{n}$ their corresponding energies. 
The optical conductivity
and Drude weight in the $y$ direction can be obtained and expressed 
similarly. $N$ is the number of sites. 
In our calculation, the Dirac $\delta$ functions are regularized as a Lorentzian
$\delta(\omega)$$\approx$$(1/\pi)\epsilon/(\omega^{2}+\epsilon^{2})$
with a small but finite broadening parameter $\epsilon$.

\begin{figure}[htb]
\begin{center}
\includegraphics[clip,width=87mm,angle=0]{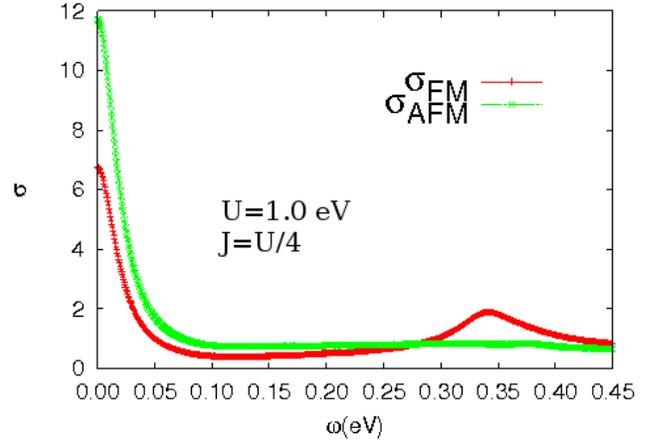}
\vskip -0.3cm
\caption{(color online) Example showing $\sigma(\omega)$
in the ``physical region''\cite{luo} of the three-orbital model ($\epsilon$=0.02). 
The unit of
 $\sigma(\omega)$ is  $e^{2}/\hbar$. 
The couplings are $U=1.0$~eV and $J$=$U/4$.
The AFM direction (i.e. the $x$ direction for magnetic wavevector $(\pi,0)$) 
has a larger zero frequency conductivity than the FM direction,
as in experiments. The FM direction also has a peak at a finite
frequency $\sim$$J$.} \label{sigma}
\vskip -0.95cm
\end{center}
\end{figure}

{\it Results.} One of the main results found in our study is shown in Fig.~\ref{sigma} where
$\sigma(\omega)$ in the two directions is shown for a state with magnetic
order ($\pi$,0). The values of the couplings $U$ and $J$ are representative 
of the so-called ``physical
region'' that was previously unveiled for the
same three-orbital model.\cite{luo} In other words, by a comparison between neutron scattering
and photoemission experiments against mean-field results, in
previous studies it was concluded
that the three-orbital model has a ``physical region'' (where theory matches
experiments) in the range $U$$\sim$[0.7,1.3] and $J/U$$\sim$[0.15,0.33],\cite{luo} where the state is
simultaneously magnetic and metallic as in pnictide parent compounds. Our
$\sigma(\omega)$ study is restricted to that ``physical region''.

\begin{figure}[htb]
\begin{center}
\includegraphics[clip,width=87mm,angle=0]{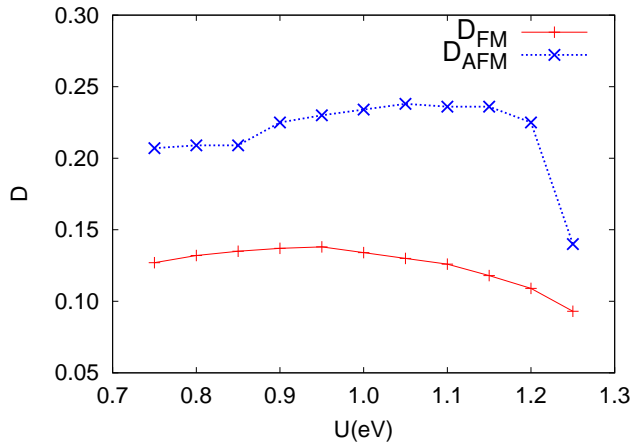}
\vskip -0.4cm
\caption{(color online) 
Drude weight/$\pi$ $vs.$ $U$ in the ``physical region'' of the three-orbital model, 
at $J$=$U/4$. 
In this regime, the inequality $D_{\rm AFM}$$>$$D_{\rm FM}$ holds. As $U$ increases toward the
upper limit shown, the Drude weights in both directions are reduced due to increasing 
insulating tendencies.\cite{luo} 
} 
\vskip -0.7cm
\label{D-figure}
\end{center}
\end{figure}

Figure~\ref{sigma} shows that
$\sigma(\omega)$ for the three-orbital model is found to be in good qualitative agreement with experiments, namely
at small frequency $\omega$, where the Drude peak is located, the weight of this peak
is larger along the AFM direction (the $x$ direction) 
than along the FM direction. Similar
results were obtained in the entire ``physical region'', see Fig.~\ref{D-figure}.
In addition, the
finite frequency peak in the FM direction was found to scale with $J$. The ratio
$D_{\rm AFM}$/$D_{\rm FM}$ (i.e. $D_x$/$D_y$) 
in the range of $U$ shown in Fig.~\ref{D-figure} varies
approximately between 1.6 and 2.2, in qualitative agreement with results for the five-orbital model.\cite{valenzuela,tohyama}
Thus, it is here concluded that the three-orbital model~\cite{three}
is sufficient to reproduce the d.c. conductivity anisotropy found in experiments.~\cite{aniso}

\begin{figure}[htb]
\begin{center}
\includegraphics[clip,width=87mm,angle=0]{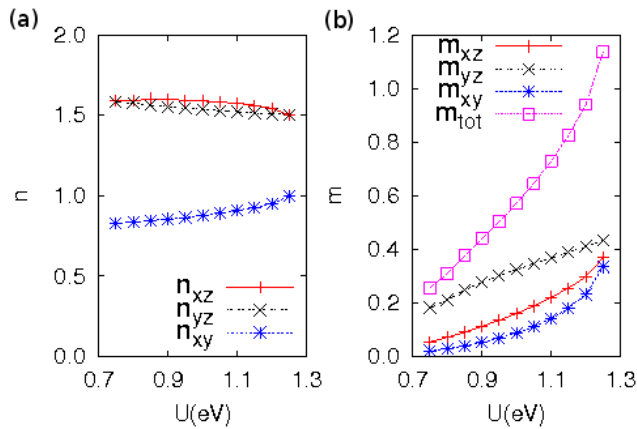}
\vskip -0.4cm
\caption{(color online) 
(a) Charge density of each orbital $vs.$ $U$ in the ``physical region'' 
of the three-orbital model, at $J$=$U/4$, and for
spin order $(\pi,0)$. 
(b) $m_{\alpha}$ $vs.$ $U$ in the same $U$ range, 
at $J$=$U/4$, and for spin order $(\pi,0)$. 
} 
\vskip -0.4cm
\label{nm-figure}
\end{center}
\end{figure}

For completeness, in Fig.~\ref{nm-figure}(a) 
the population of the three orbitals is shown
in the range of $U$'s studied. From this figure, it is clear that there is no 
orbital order since the orbitals $d_{xz}$ and $d_{yz}$ are nearly identically populated.
Further increasing $U$ eventually leads to a regime of orbital-order,\cite{three} 
but the opening of a gap renders the system insulating. It is important
to note that Fig.~\ref{nm-figure}(a) contains results obtained by
integrating the orbital-selective density-of-states over all frequencies, while if the focus is only
the vicinity of the FS, the orbital-weight redistribution phenomenon is observed.\cite{three}
As shown below, this redistribution is important to understand the anisotropy.

In addition, Fig.~\ref{nm-figure}(b) shows 
that the magnetic moment in the range
investigated is compatible with pnictides
neutron experiments, 
ranging from $\sim$0.25 Bohr magnetons ($\mu_B$)
for the 1111 to $\sim$1~$\mu_B$ for the 122 compounds.\cite{luo}

{\it Intuitive origin of the anisotropy.} While the notion of an orbital weight redistribution
at the FS is well established,\cite{owr,valenzuela} with the $d_{yz}$ orbital suppressed for
${\bf Q}$=($\pi$,0), it is desirable to develop a more intuitive understanding of its influence
on transport properties. For this purpose, 
the kinetic energy and current operators will be expressed 
in momentum space as:

\begin{equation}
\begin{split}
 {\hat{T}_a}&=\sum_{\bf{k}}\sum_{\alpha, \beta, \sigma}t_{a}^{\alpha \beta}({\bf{k}})c^{\dagger}_{{\bf{k}},\alpha,\sigma}c_{{\bf{k}},\beta,\sigma}=\sum_{\bf{k}}{\hat{T}_a}({\bf{k}}),\\
 {\hat{j}_a}&=\sum_{\bf{k}}\sum_{\alpha, \beta, \sigma}j_{a}^{\alpha \beta}({\bf{k}})c^{\dagger}_{{\bf{k}},\alpha,\sigma}c_{{\bf{k}},\beta,\sigma}=\sum_{\bf{k}}{\hat{j}_a}({\bf{k}}),\\
\end{split}
\end{equation}


\noindent where $a$ is the direction index ($x$, $y$). 
From Eq.(4), the ${\bf k}$ contribution 
(unfolded first Brillouin zone) to the Drude weight
is defined as $D_a({\bf{k}})$=$D_{1,a}({\bf{k}})$-$D_{2,a}({\bf{k}})$,
where
\begin{equation}\begin{split}
 \frac{D_{1,a}({\bf{k}})}{2\pi}&=\frac{\langle \phi_{0}|-{\hat{T}_a}({\bf{k}})|\phi_{0} \rangle}{2N},\\
 \frac{D_{2,a}({\bf{k}})}{2\pi}&=\frac{1}{N}\sum_{n\not=0}    
 Re\frac{\langle \phi_{0}|{\hat{j}_a}({\bf{k}})|\phi_{n} \rangle \times \langle \phi_{n}|{\hat{j}_a}|\phi_{0} \rangle}
{E_{n}-E_{0}},\\
\end{split}\end{equation} 
since summing over $\bf{k}$  leads to Eq.~(4).

\begin{figure}[htb]
\begin{center}
\includegraphics[clip,width=90mm,angle=0]{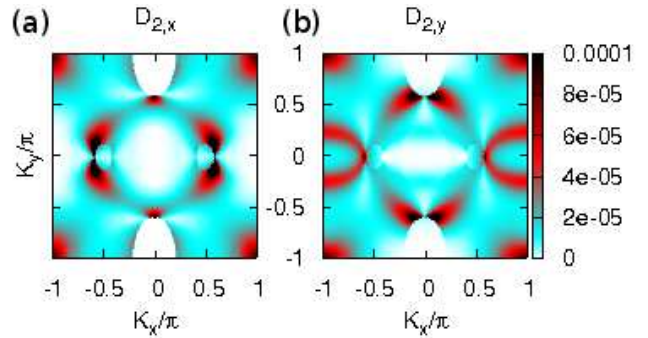}
\caption{(color online) 
$D_{2,a}(\bf{k})/\pi$ in the ``physical region''\cite{luo} 
of the three-orbital model ($U$=$1.0$~eV, $J$=$U/4$) 
and with spin order $(\pi,0)$. For a discussion of the results see text.
} 
\vskip -0.4cm 
\label{Drudek}
\end{center}
\end{figure}

From Fig.~\ref{Drudek}(b) it is clear that the $(\pi,0)$ pocket contributes
significantly to $D_{2,y}$ since this wavevector region has sizable intensity. However,  
the contribution of the $(0,\pi)$ pocket to $D_{2,x}$ is negligible, thus inducing the
significant anisotropy observed in the overall Drude weight (note that the rest of the highly intense
features in $D_{2,y}$ are $\pi$/2-rotated those of $D_{2,x}$ and thus do not contribute to the anisotropy).
The reason is the orbital weight redistribution at the FS:
according to nesting scenarios, the $(0,0)$ pocket (with mainly $d_{xz}$
and $d_{yz}$ character), interacts with the 
$(\pi,0)$ pocket (mainly  $d_{xy}$ and $d_{yz}$) when the magnetic wavevector 
is $\bf{Q}$=($\pi$,0). This
interaction needs to be intra-orbital,\cite{recent} hence the $d_{yz}$ states at the $(\pi,0)$ pocket 
are raised above the FS, while the $d_{xy}$ states are not. 
On the other hand, the $(0,\pi)$ states are not 
moved above the FS if ${\bf Q}$=($\pi$,0), 
and hence this phenomenon does not occur for $D_{2,x}$. Note that the results of Fig.~\ref{Drudek} 
based on the Drude weights directly address the anisotropy in transport, and complements 
the analysis based on the orbital-weight redistribution.\cite{owr,valenzuela}
Also note that a similar analysis of $D_{1,y}$ and $D_{1,x}$ (not shown) does not lead to the same
clear anisotropy that $D_{2,a}$ provides.

To further simplify the understanding of the anisotropy 
evident in Fig.~\ref{Drudek} let us now focus
on the most relevant electronic hopping processes. 
Analyzing the values of the hopping amplitudes (Table~1), 
it is clear that those involving the $d_{xy}$ orbital (both
inter- and intra-orbital) should be the most relevant since the other hoppings
are much smaller in magnitude. The direct 
intra-orbital hopping $d_{xy}$-$d_{xy}$ ($t^{33}_{\bf{il}}$ in Table~1) 
is not suppressed at the FS and should equally
contribute to charge transport in both directions. Thus,
the conductance should not drop to zero in any of the two directions 
due to this intra-orbital contribution. However,
the inter-orbital hopping $d_{xy}$-$d_{yz}$ is suppressed 
at the FS in a magnetic state ($\pi$,0).
This hopping occurs {\it only} along the $y$ direction, 
as previously discussed.\cite{three}
On the other hand, the hopping $d_{xy}$-$d_{xz}$ is {\it not} 
suppressed and can contribute
to electronic hopping along the $x$ direction. 
For these reasons, an asymmetry is expected between 
the $x$ and $y$ directions in transport, as found in
Fig.~\ref{sigma}. In addition, for $\bf{k}$ close to the $(\pi,0)$ 
pocket the inter-orbital $d_{xy}$-$d_{yz}$ hopping, 
which only exists along the $y$ direction, 
needs an excitation to contribute to $\sigma(\omega)$ 
along the FM direction ($y$-axis) because
$d_{yz}$ is suppressed at the FS. This observation justifies 
the presence of a peak scaling with $J$ in $\sigma_{\rm FM}$ (Fig.~1).

To transform the intuition developed above based on hopping amplitudes into actual transport
properties, the Drude weight will also be decomposed according to the
hoppings corresponding to the different orbitals, via the following 
definitions:

\begin{equation}\begin{split}
 \frac{D_{a}^{\alpha \beta}}{2\pi}&=\frac{\langle \phi_{0}|-{\hat{T}}_{a}^{\alpha \beta}-{\hat{T}}_{a}^{\beta \alpha}|\phi_{0} \rangle}{2N}\\
                              &-\frac{1}{N}\sum_{n\not=0}
    Re\frac{\langle \phi_{0}|{\hat{j}}_{a}^{\alpha \beta}+{\hat{j}}_{a}^{\beta \alpha}|\phi_{n} \rangle\langle \phi_{n}|{\hat{j}_a}|\phi_{0} \rangle}{E_{n}-E_{0}},\\
\end{split}\end{equation}
for inter-orbital hopping ($\alpha$$\neq$$\beta$) ($a$=$x$,$y$).
The operators in Eq.~(7) arise from Eq.~(2) via
${\hat{T}}_a$=$\sum_{a,\alpha \beta} {\hat{T}}_{a}^{\alpha \beta}$ and
${\hat{j}}_a$=$\sum_{a,\alpha \beta} {\hat{j}}_{a}^{\alpha \beta}$.
For the case of intra-orbital, the diagonal
Drude weight $D_{a}^{\alpha \alpha}$ is obtained from Eq.~(7) by replacing
${\hat{T}}_{a}^{\alpha \beta}$+${\hat{T}}_{a}^{\beta \alpha}$ 
by ${\hat{T}}_{a}^{\alpha \alpha}$ and
${\hat{j}}_{a}^{\alpha \beta}$+${\hat{j}}_{a}^{\beta \alpha}$ 
by ${\hat{j}}_{a}^{\alpha \alpha}$.
The several Drude components obtained by this procedure 
are in Table~\ref{tab:Dab3}. From this Table, it can be seen 
that the main anisotropy arises from the fact that 
$D_{x}^{13}$$\sim$0.125 is an order of magnitude 
larger than $D_{y}^{23}$$\sim$0.011, due to the 
FS suppression of the $d_{yz}$ orbital. The rest of the 
contributions in Table~\ref{tab:Dab3} that are unrelated to the
inter-orbital hopping involving $d_{xy}$ are similar 
in both directions and are not
relevant to understand the anisotropy. Actually, for the largest of 
those, the naive intuition suggesting a better conductance 
along the FM direction {\it is} satisfied since $D_{y}^{33}$$>$$D_{x}^{33}$.


{\it Summary.} A mean-field study of $\sigma(\omega)$ employing 
a three-orbital Hubbard model for the magnetically ordered parent 
compounds of the pnictides has been here reported.
In agreement with experiments, the conductance along the AFM direction is
shown to be larger than along the FM direction. The simplicity of this model 
allowed us to reduce this effect to an intuitive picture:
(i) The AFM conductance behaves normally with
a notorious suppression of its value as compared with the non-interacting
limit due to spin scattering, in agreement with intuition. 
(ii) However, along the FM direction the drastic 
reduction in the weight of the $d_{yz}$
orbital at the FS leads to a large effective {\it suppression} 
of the $d_{xy}$-$d_{yz}$ hopping and associated conductance along 
that FM direction, causing the anisotropy found experimentally.

\vskip 0.7cm
\begin{table}
\vspace{0.1cm}
 \centering
 \begin{tabular}{|p{0.9cm}|p{1.2cm}p{1.2cm}p{1.2cm}p{1.2cm}p{1.2cm}p{0.9cm}|}\hline
$\alpha\beta$ &$11$       &$22$       &$33$             &$12$      &$13$        &$23$ \\
\hline
$D_{x}^{\alpha \beta}$ &$0.019$   &$-0.003$     &$0.073$   &$0.013$  & $0.125$     & $0.005$     \\
\hline
$D_{y}^{\alpha \beta}$ &$0.020$   &$0.002$      &$0.087$   &$0.014$  & $-0.01$     & $0.011$      \\
\hline
 \end{tabular}
\caption{Drude weight$/\pi$ decomposed into the different orbitals 
($1$=$d_{xz}$, $2$=$d_{yz}$, $3$=$d_{xy}$) of the  three-orbital model working
 at $U$=$1.0$~eV and $J$=$0.25U$.
 Finding negative Drude weights in some cases is a well-known
effect\cite{Elbio} arising from differences of two large numbers in Eq.~(4).
\label{tab:Dab3}} 
\vskip -0.5cm
\end{table}

{\it Acknowledgments.} Work supported by 
the U.S. Department of Energy, Office of Basic Energy Sciences,
Materials Sciences and Engineering Division.

\end{document}